\documentclass[aps,prl,twocolumn,showpacs,preprintnumbers,superscriptaddress]{revtex4-1}

\usepackage{amsmath,amssymb,amsfonts}
\usepackage{graphicx}
\usepackage{dsfont}

\def\tr{\mathrm{Tr}}
\def\re{\mathrm{Re}}
\def\({\left(}
\def\){\right)}
\def\I{\mathcal{I}}
\def\N{\mathcal{N}}
\newcommand{\lettersection}[1]{\noindent{\bf #1 ---}}

\begin{document}

\preprint{CERN-PH-TH-2014-186}

\title{Lattice gauge theory without link variables}

\author{H\'{e}lvio Vairinhos}
\affiliation{Institute for Theoretical Physics, ETH Z\"urich, CH-8093 Z\"urich, Switzerland}
\affiliation{CFP, Department of Physics, University of Porto, 4169-007 Porto, Portugal}

\author{Philippe de Forcrand}
\affiliation{Institute for Theoretical Physics, ETH Z\"urich, CH-8093 Z\"urich, Switzerland}
\affiliation{CERN, Physics Department, TH Unit, CH-1211 Geneva 23, Switzerland}

\date{September 30, 2014}

\begin{abstract}
We obtain a sequence of alternative representations for the partition function of pure $SU(N)$ or $U(N)$ lattice gauge theory with the Wilson plaquette action, using the method of Hubbard-Stratonovich transformations. In particular, we are able to integrate out all the link variables exactly, and recast the partition function of lattice gauge theory as a Gaussian integral over auxiliary fields.
\end{abstract}

\pacs{11.15.Ha, 12.38.Aw, 12.38.Gc}

\maketitle

\lettersection{Introduction} One of the most ambitious programs in lattice gauge theory is to map the phase diagram of QCD at finite temperature and density from first principles. The difficulty of this program resides in the fact that non-zero chemical potentials generally imply complex-valued fermionic actions. This leads to a severe sign problem that prevents the direct sampling of the grand canonical ensemble of lattice QCD using standard Monte Carlo techniques \cite{deForcrand:2010ys}.

A promising approach to tackle this sign problem is the worldline representation of lattice QCD \cite{Rossi:1984cv}, where link variables are integrated out {\em before} the (staggered) fermions. This contrasts with the traditional method of integrating out the Grassmann variables first. 

The heuristic argument for the worldline approach is that large cancellations in the path integral of finite density QCD are driven by gauge fluctuations, hence by integrating out the gauge degrees of freedom first we hope that the resulting sign problem becomes milder.

A limitation of the worldline approach is that exact integration of the link variables is only known to be possible in the strong coupling limit, $\beta = 0$. In this limit, the plaquette terms drop from the lattice action, and group integration over the link variables reduces to a product of solvable fermionic one-link integrals \cite{Rossi:1984cv}. After the gauge integration at $\beta=0$, the remaining degrees of freedom are worldlines of free color singlets. 

Subsequently, after integrating out all the Grassmann variables, and after a clever resummation of the final result \cite{Karsch:1988zx}, the partition function of the strong coupling limit of lattice QCD reduces to a rather simple monomer-dimer-polymer (MDP) system. 

Recent simulations of this model \cite{Fromm:2008ab,*deForcrand:2009dh,*Fromm:2010lga,*Unger:2012jt} and of its $O(\beta)$ corrections \cite{Fromm:2011kq,*deForcrand:2013ufa} using worm-inspired algorithms \cite{Prokofiev:2001,Adams:2003cca,*Chandrasekharan:2006tz}, have allowed to map the whole phase diagram of strong coupling lattice QCD, and to confirm that the sign problem in the MDP model is mild enough to be tractable with reweighting methods. However, going beyond the $O(\beta)$ corrections in the strong coupling expansion of the MDP model leads to rather cumbersome expressions. 

In order to approach the regime of continuum physics, it would be desirable to have a simpler MDP model of lattice QCD for arbitrary values of the lattice coupling. However, this would require evaluating unitary group integrals in the presence of plaquette terms, which cannot be done directly with the available mathematical tools. 

As a first step in this direction, we show in this Letter how to integrate out exactly all the link variables in the canonical partition function of pure lattice gauge theory with a unitary gauge group and the Wilson plaquette action, for any value of the lattice coupling.

Our method consists of replacing the unitary group integrals over the link variables with Gaussian integrals over a set of auxiliary variables, using suitable Hubbard-Stratonovich transformations. Then, the Gaussian integrals over the auxiliary variables may either be solved exactly in the simplest cases, or directly sampled with simple heatbath algorithms \citep{Box:1958}.

Trading the original link variables for auxiliary Gaussian variables achieves a decoupling of the four links originally coupled around a plaquette. In turn, this allows the 1-link integrals to be performed analytically, even in the presence of quark fields, for any value of the plaquette coupling $\beta$. We discuss the promise of this approach in the conclusion of this Letter.
\\

\lettersection{4-link action} Let us consider pure Yang-Mills theory regularized on a periodic $d$-dimensional Euclidean hypercubic lattice, with the Wilson plaquette action:
\begin{equation}
\label{eq:4-link:S}
	S_4 = 
	\beta
	\sum_{x}
	\sum_{\mu<\nu}^d 
	\(1 - \frac{1}{N} \re\tr(U_{x,\mu\nu}) \)
\end{equation}
where $U_{x,\mu\nu} \equiv U_{x,\mu}^{} U_{x+\hat\mu,\nu}^{} U_{x+\hat\nu,\mu}^\dag U_{x,\nu}^\dag$ is the plaquette matrix, $U_{x,\mu} \in SU(N)$ or $U(N)$ are the link variables, and $\beta$ is the lattice coupling. The subscript in $S_4$ serves to indicate that each term in the action contains a product of four link variables. We call it {\em 4-link action}.

The partition function of this theory is:
\begin{equation}
\label{eq:4-link:Z}
	Z = \int [dU]~ e^{-S_4}
\end{equation}
where $[dU] \equiv \prod_{x,\mu} dU_{x,\mu}$ is a product of Haar measures.

\lettersection{Gaussian measures} Let $X$ be a random complex-valued $N \times N$ matrix whose elements $X_{ij}$ are normally distributed according to a Gaussian measure of the form:
\begin{equation}
\label{eq:gauss:measure}
	\gamma_a(X) = 
	\prod_{i,j=1}^N \frac{a}{2\pi}~
	dX_{ij}^{} dX_{ij}^\ast~
	e^{-\frac{a}{2} |X_{ij}|^2}
\end{equation}
where $a>0$ is a constant. The distribution above is normalized, i.e. $\int \gamma_a(X) = 1, \, \forall a$. For $a=1$ we drop the subscript, i.e. $\gamma(X) \equiv \gamma_1(X)$.

In our notation, the composition of a Gaussian measure with a Gaussian weight gives:
\begin{equation}
\label{eq:gauss:composition}
	\gamma_a(X)~ e^{-\frac{b}{2}\tr(X^\dag X)} =
	\gamma_{a+b}(X) \left( \frac{a}{a+b} \right)^{N^2}
\end{equation}

A change of variables in the form of a linear shift:
\begin{equation}
\label{eq:gauss:HS:change-vars}
	X' = \sqrt a~ (X - Y)
\end{equation}
for constant $Y$ and $a>0$, implies the relation:
\begin{equation}
\label{eq:gauss:HS:change-meas}
	\gamma(X')~ e^{\frac{a}{2} \tr(Y^\dag Y)} = 
	\gamma_a(X)~ e^{a \re\tr(X^\dag Y)}
\end{equation}
Integrating the expression above, we get:
\begin{equation}
\label{eq:gauss:HS}
	e^{\frac{a}{2} \tr(Y^\dag Y)} = 
	\int \gamma_a(X)~ e^{a \re\tr(X^\dag Y)}
\end{equation}
which is an example of a Hubbard-Stratonovich (HS) transformation \cite{Hubbard:1959}.
\\

\lettersection{2-link action} The 4-link action \eqref{eq:4-link:S} can be expressed as a ``sum of squares":
\begin{equation} \label{eq:4-link:S:square}
	S_4 =
	-\frac{\beta}{2N}
	\sum_x
	\sum_{\mu<\nu}^d 
		\tr(W^\dag W)_{x,\mu\nu} 
	+2\beta N_P
\end{equation}
where $N_P = \frac{1}{2}d(d-1) V$ is the total number of plaquettes, $V$ being the lattice volume. $W_{x,\mu\nu}$ is the complex-valued $N \times N$ matrix defined by:
\begin{equation}
\label{eq:link:diagonal}
	W_{x,\mu\nu} = W_{x,\nu\mu} = 
        U_{x,\mu} U_{x+\hat\mu,\nu} + U_{x,\nu} U_{x+\hat\nu,\mu}
\end{equation}
which can be thought of as a ``square root" of a plaquette.

\begin{figure}
\includegraphics[scale=0.8]{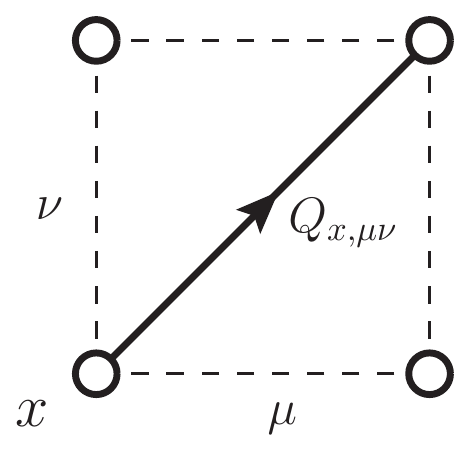}
\qquad
\includegraphics[scale=0.8]{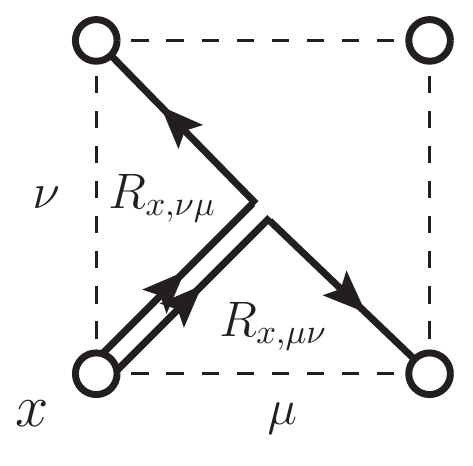}
\caption{
\label{fig:auxfields}
Graphical representation of the auxiliary variables necessary for the construction of the $n$-link actions. The diagonal link (left) splits the original plaquette into two halves ($\frac{1}{2}$-plaquettes), while the folded links (right) split each half into two quarters ($\frac{1}{4}$-plaquettes).
}
\end{figure}

Let $Q_{x,\mu\nu}'$ ($= Q_{x,\nu\mu}')$ be random complex-valued $N \times N$ matrices with normal distribution $\gamma(Q_{x,\mu\nu}')$; they are naturally associated with the ``diagonal link'' connecting the lattice sites $x$ and $x + \hat\mu + \hat\nu$ (see Fig.\ref{fig:auxfields}).

Using the HS transformation \eqref{eq:gauss:HS} for a change of variables of the form:
\begin{equation}
\label{eq:2-link:HS}
	Q'_{x,\mu\nu} = \sqrt{\frac{\beta}{N}}~
	\( Q_{x,\mu\nu} - W_{x,\mu\nu} \)
\end{equation}
the Boltzmann weight of the partition function \eqref{eq:4-link:Z} can be expressed as a Gaussian integral over diagonal links:
\begin{eqnarray}
\label{eq:4-link:Boltzmann}
	e^{-S_4}
	&=&
	e^{-2\beta N_P} \!\!
	\prod_{x,\mu<\nu}
	e^{\frac{\beta}{2N}
		\tr(W^\dag W)_{x,\mu\nu}
	}
	\nonumber \\
	&=& 
	e^{-2\beta N_P} \!\!
	\prod_{x,\mu<\nu}
	\int \gamma_{\frac{\beta}{N}}(Q_{x,\mu\nu})~ 
	e^{\frac{\beta}{N}\re\tr(Q^\dag W)_{x,\mu\nu}}
	\nonumber \\
	&=&
	\int \gamma_{\frac{\beta}{N}}[Q]~ e^{-S_2}
\end{eqnarray}
where $\gamma_{\frac{\beta}{N}}[Q] \equiv \prod_{x,\mu<\nu} \gamma_{\frac{\beta}{N}}(Q_{x,\mu\nu})$ is a product of Gaussian measures, and $S_2$ is the {\em 2-link action}:
\begin{equation}
\label{eq:2-link:S}
	S_2 = \beta
		\sum_{x} 
		\sum_{\mu\neq\nu}^d 
		\(1-\frac{1}{N} \re\tr\(
		Q_{x,\mu\nu}^\dag 
		U_{x,\mu}^{}
		U_{x+\hat\mu,\nu}^{}
		\)\)
\end{equation}
The partition function \eqref{eq:4-link:Z} then becomes:
\begin{equation}
\label{eq:2-link:Z}
	Z = 
	\int \gamma_{\frac{\beta}{N}}[Q] 
	\int [dU]~ e^{-S_2}
\end{equation}
Graphically, each term of the 2-link action represents the contribution of a ``$\frac{1}{2}$-plaquette'' composed of one diagonal link and two ordinary links.

The method of splitting plaquette terms of the Wilson action into $\frac{1}{2}$-plaquette terms was originally proposed by Fabricius and Haan in the context of the Twisted Eguchi-Kawai model \cite{Fabricius:1984wp}. It was also used for lattice simulations of non-commutative $U(1)$ gauge theory \cite{Bietenholz:2006cz}, and later extended to certain classes of lattice gauge actions that have a polynomial dependence on the link variables \cite{Vairinhos:2010ha}.
\\

\lettersection{1-link action} The 2-link action \eqref{eq:2-link:S} also can be expressed as a ``sum of squares":
\begin{eqnarray}
\label{eq:2-link:S:square}
	S_2 
	&=&
	-\frac{\beta}{2N}
	\sum_{x} 
	\sum_{\mu\neq\nu}^d 
	\tr(W^\dag W)_{x,\mu\nu}
	\nonumber \\
	&&
	+
	\frac{\beta}{N} 
	\sum_{x}
	\sum_{\mu<\nu}^d 
	\tr(Q^\dag Q)_{x,\mu\nu}
	+ 3\beta N_P
\end{eqnarray}
where $W_{x,\mu\nu}$ is now defined by:
\begin{equation}
\label{eq:wedge-link}
	W_{x,\mu\nu} = 
	Q_{x,\mu\nu}^{} 
	U_{x+\hat\mu,\nu}^\dag + 
	U_{x,\mu}^{}
\end{equation}

Let $R_{x,\mu\nu}'$ be random complex-valued $N \times N$ matrices with normal distribution $\gamma(R_{x,\mu\nu}')$. $R_{x,\mu\nu}'$ $(\neq R_{x,\nu\mu}')$ is naturally associated with the ``folded link'' connecting the lattice sites $x$ and $x + \hat\mu$ and contained in the $(\mu,\nu)$-plaquette (see Fig.\ref{fig:auxfields}).

Using the HS transformation \eqref{eq:gauss:HS} for a change of variables of the form:
\begin{equation}
\label{eq:1-link:HS}
	R_{x,\mu\nu}' = 
	\sqrt{\frac{\beta}{N}}~
	\( R_{x,\mu\nu} - W_{x,\mu\nu} \)
\end{equation}
the Boltzmann weight of the partition function \eqref{eq:2-link:Z} can be expressed as a Gaussian integral over folded links:
\begin{eqnarray}
\label{eq:2-link:Boltzmann}
	e^{-S_2}
	&=&
	\N
	e^{-\frac{\beta}{N} \tr[Q^\dag Q]} \!\!
	\prod_{x,\mu\neq\nu}
	e^{\frac{\beta}{2N} \tr(W^\dag W)_{x,\mu\nu}}
	\nonumber \\
	&=& 
	\N
	e^{-\frac{\beta}{N} \tr[Q^\dag Q]} \!\!
	\prod_{x,\mu\neq\nu}
	\int \gamma_{\frac{\beta}{N}}(R_{x,\mu\nu})~
	e^{\frac{\beta}{N} \re\tr(R^\dag W)_{x,\mu\nu}}
	\nonumber \\
	&=&
	\N
	e^{-\frac{\beta}{N} \tr[Q^\dag Q]}
	\int \gamma_{\frac{\beta}{N}}[R]~ e^{-S_1}
\end{eqnarray}
where $\N = e^{-3\beta N_P}$ is a normalization factor, $\gamma_{\frac{\beta}{N}}[R] \equiv \prod_{x,\mu\neq\nu} \gamma_{\frac{\beta}{N}}(R_{x,\mu\nu})$ is a product of Gaussian measures, $\tr[Q^\dag Q] \equiv \sum_{x,\mu<\nu} \tr(Q^\dag Q)_{x,\mu\nu}$ is a contribution from diagonal links, 
and $S_1$ is the {\em 1-link action}:
\begin{equation}
\label{eq:1-link:S}
	S_1 = 
	-\frac{\beta}{N} \sum_{x,\mu} 
	\re\tr\(J_{x,\mu}^\dag U_{x,\mu}^{}\)
\end{equation}
where $J_{x,\mu}$ depends on the auxiliary variables only:
\begin{equation}
\label{eq:J}
	J_{x,\mu} = \sum_{\nu=1 \atop (\nu\neq\mu)}^d
	(
		R_{x-\hat\nu,\nu\mu}^\dag Q_{x-\hat\nu,\mu\nu}^{}
		+ R_{x,\mu\nu}^{}
	)
\end{equation}

Graphically, each term of the 1-link action represents the contribution of a ``$\frac{1}{4}$-plaquette'' composed of one folded link and one ordinary link (or of one folded, one diagonal and one ordinary link, which effectively covers a different $\frac{1}{4}$-plaquette).

Using \eqref{eq:gauss:composition}, we get:
\begin{equation}
\label{eq:1-link:diag-meas}
	\gamma_{\frac{\beta}{N}}(Q_{x,\mu\nu})~
	 e^{-\frac{\beta}{N} \tr(Q^\dag Q)_{x,\mu\nu}^{}} = 
	 \gamma_{\frac{3\beta}{N}}(Q_{x,\mu\nu})~ 3^{-N^2}
\end{equation}
and the partition function becomes:
\begin{equation}
\label{eq:1-link:Z}
	Z = \N_1 \int
	\gamma_{\frac{3\beta}{N}}[Q] 
	\gamma_{\frac{\beta}{N}}[R] \int [dU]~ e^{-S_1}
\end{equation}
where $\N_1 = e^ {-(3\beta + N^2 \log 3) N_P}$. 
\\

\lettersection{0-link action} The partition function \eqref{eq:1-link:Z} is a multiple Gaussian integral whose integrand clearly factorizes as a product of one-link integrals, also known as Br\'{e}zin-Gross-Witten (BGW) integrals: 
\begin{equation}
\label{eq:group_integral}
	{\cal I}_G(J,J^\dag) = 
	\int_G dU~ e^{\tr(J U^\dag + U J^\dag)}
\end{equation}
where $G=SU(N)$ or $U(N)$ is the gauge group, and $J$ is a complex $N \times N$ matrix.

Exact solutions of BGW integrals for general $J$ are known in closed form for some unitary groups of small rank \cite{Eriksson:1980rq}. Each of those solutions provides an alternative representation, without link variables, of the partition function of the corresponding lattice gauge theory:
\begin{eqnarray}
\label{eq:0-link:Z}
	Z &=& 
	\N_0
	\int 
	\gamma_{\frac{3\beta}{N}}[Q] 
	\gamma_{\frac{\beta}{N}}[R]~ 
	\prod_l {\cal I}_G\! \(
	\frac{\beta}{2N} J_l^{},
	\frac{\beta}{2N} J_l^\dag
	\)
	\nonumber\\
	&=&
	\N_0
	\int 
	\gamma_{\frac{3\beta}{N}}[Q] 
	\gamma_{\frac{\beta}{N}}[R]~ 
	e^{-S_0}
\end{eqnarray}
where $l$ labels lattice links, $S_0$ is the {\em 0-link action}:
\begin{equation}
\label{eq:0-link:S}
	S_0 = -\sum_l 
	\log \I_G\! \(\frac{\beta}{2N} J_l^{}, \frac{\beta}{2N} J_l^\dag\)
\end{equation}
and the normalization factor is $\mathcal N_0 = e^ {-(3\beta + N^2 \log 3) N_P}$.

In the $SU(2)$ case, for example, the one-link integral is very simple \cite{Eriksson:1980rq} and the 0-link action reduces to:
\begin{equation}
\label{eq:0-link:S:SU2}
	S_0 = 
	-\sum_l 
	\log \( 
	\frac{2 I_1(\frac{\beta}{2} z_l)}{\frac{\beta}{2} z_l}
	\)
\end{equation}
where $z_l^2 = \tr(J_l^{} J_l^\dag) + \det(J_l) + \det(J_l^\dag)$ is an $SU(2)$ invariant, and $I_1(z)$ is a modified Bessel function of the first kind.

The auxiliary variables which we have introduced transform covariantly under a local gauge transformation, so that our expressions for the actions are naturally gauge-invariant. Center symmetry is preserved as well.
\\

\lettersection{Observables} For $n\geq 1$, gauge-invariant observables retain their original definition in terms of link variables. The auxiliary fields decouple from the link variables after an inverse HS transformation, so the expectation values of lattice observables must not depend on them. Only statistical fluctuations are affected, which can be seen in Table \ref{tab:plaquettes}.

However, for $n=0$ the link variables are integrated out. In this case, bulk observables (e.g. energy density, specific heat, etc.) can be obtained from derivatives of the 0-link partition function with respect to $\beta$. 
For example, we may define the energy density by:
\begin{eqnarray}
\label{eq:energy-density}
	\!\!\varepsilon(\beta)
	&=& -\frac{1}{N_P} \frac{\partial}{\partial\beta} \log Z
	= 1-\langle u_p \rangle
	\nonumber \\
	&=& 
	3 + \frac{3}{2\beta N N_P}\! \sum_{x,\mu<\nu} \!
	\left\langle \beta
	\tr(Q_{x,\mu\nu}^\dag Q_{x,\mu\nu}^{}) 
	\right\rangle
	\nonumber \\
	&& + 
	\frac{1}{2\beta N N_P}\! \sum_{x,\mu\neq\nu} \!
	\left\langle \beta
	\tr(R_{x,\mu\nu}^\dag R_{x,\mu\nu}^{}) 
	\right\rangle
	- \frac{3N^2}{\beta}
	\nonumber \\
	&& - 
	\frac{1}{N_P} \sum_l 
	\left\langle
	\frac{\partial}{\partial\beta}
	\log
	\mathcal{I}_G\!\(
		\frac{\beta}{2N} J_l^{},
		\frac{\beta}{2N} J_l^\dag
	\)\!\!
	\right\rangle
\end{eqnarray}
where $u_p$ is the plaquette operator.
The first term in the r.h.s. of the expression above is the contribution from the normalization constant, the next three terms are contributions from the Gaussian measure, and the last term is the contribution from the 0-link action \eqref{eq:0-link:S}. 
In particular for $SU(2)$, the contribution from the 0-link action reduces to:
\begin{eqnarray}
\label{eq:energy-density:SU2}
	\frac{\partial}{\partial\beta}
	\log
	\mathcal{I}_G\!\(
		\frac{\beta}{2N} J_l^{},
		\frac{\beta}{2N} J_l^\dag
	\)
	=
		\frac{z_l}{2}
		\frac{I_2(\frac{\beta}{2} z_l)}{I_1(\frac{\beta}{2} z_l)}
		~~
\end{eqnarray}

The apparently divergent contributions coming from the Gaussian measure cancel out, and result in a finite quantity that vanishes at $\beta=0$. However, the cancellations are difficult to control during Monte Carlo simulations at very strong coupling. In that situation it is natural to expect increased variance in bulk observables.

Other observables require a re-evaluation of the group integrals, to take into account the link variables in their definition. This can also be achieved by taking derivatives of \eqref{eq:group_integral} with respect to the sources $J_l$. 
For example, the expectation value of the Wilson loop operator over a non-self-intersecting closed curve $C$ is given by:
\begin{eqnarray}
\label{eq:WilsonLoop}
	\langle W(C) \rangle
	&=& 
	\frac{\N_1}{Z} \!
	\int\!
	\gamma_{\frac{3\beta}{N}}[Q] 
	\gamma_{\frac{\beta}{N}}[R]
	\!\int [dU] e^{-S_1}
	\frac{1}{N}
	\tr\!\(\! \mathcal{P} \prod_{l\in C} U_l \!\)
	\nonumber \\
	&=&
	\left\langle
		\frac{1}{N} \tr\(\! \mathcal{P} \prod_{l\in C} \widetilde U_l \!\)\!\!
	\right\rangle
\end{eqnarray}
where products are path-ordered around $C$, and $\widetilde U_l$ is the ``effective link'' defined by:
\begin{eqnarray}
\label{eq:effective-link}
	\widetilde U_l^{ij}
	&=&
		\frac{1}{\mathcal{I}_G(\frac{\beta}{2N} J_l, \frac{\beta}{2N} J_l^\dag)}
		\int_G dU~ e^{\frac{\beta}{N} \re\tr(J_l^\dag U)}~ U^{ij}
	\nonumber\\
	&=&
		\frac{2N}{\beta}
		\frac{\partial}{\partial (J_l^\dag)^{ji}}
		\log \I_G \!\(\frac{\beta}{2N} J_l^{}, \frac{\beta}{2N} J_l^\dag\)
\end{eqnarray}
In particular, the effective link for $SU(2)$ is given by:
\begin{equation}
\label{eq:effective-link:SU2}
	\widetilde U_l^{}
	= \frac{1}{z_l} 
	\frac{I_2(\frac{\beta}{2} z_l)}{I_1(\frac{\beta}{2} z_l)} 
	(J_l^{} + {\rm adj}(J_l^\dag))
\end{equation}
where ${\rm adj}(J_l^\dag)$ is the adjugate matrix of $J_l^\dag$.

Polyakov loops are defined in the same way. From the second line of \eqref{eq:effective-link} it is clear that they are covariant but not invariant under the global center symmetry now applied to $J_l$, which still makes them suitable order parameters for its spontaneous breaking.
\\

\lettersection{Monte Carlo simulations} In numerical simulations of $n$-link actions ($n \geq 1$), link and auxiliary variables are treated on an equal footing when it comes to local updates. In practice, diagonal and folded links are updated with a Gaussian heatbath \cite{Box:1958}, followed by the HS transformations \eqref{eq:2-link:HS} and \eqref{eq:1-link:HS}, respectively; the unitary link variables are updated with the Cabibbo-Marinari pseudo-heatbath algorithm \cite{Cabibbo:1982zn}, taking into account their coupling to all surrounding links (ordinary, diagonal and folded).

We have simulated the $n$-link actions numerically, and compared the expectation values of the plaquette operator for fixed values of the lattice parameters. They coincide within statistical errors, as expected (see Table \ref{tab:plaquettes}).

\begin{table}
\begin{ruledtabular}
\begin{tabular}{lcccccc}
        & \multicolumn{2}{c}{$U(1)$}
        & \multicolumn{2}{c}{$SU(2)$}
        & \multicolumn{2}{c}{$SU(3)$} 
        \\
        & $\beta$ & $\langle u_p \rangle$
        & $\beta$ & $\langle u_p \rangle$
        & $\beta$ & $\langle u_p \rangle$
        \\ \hline
$S_{4}$              & 1.00 & 0.58529(20) & 2.25 & 0.586199(19) & 5.70 & 0.549189(18) \\
$S_{2}$              & 1.00 & 0.58526(37) & 2.25 & 0.586240(29) & 5.70 & 0.549218(39) \\
$S_{1}$              & 1.00 & 0.58556(55) & 2.25 & 0.586247(53) & 5.70 & 0.549068(64)   \\
${S_{0}}^\dag$              & 1.00 & 0.58555(55) & 2.25 & 0.586252(53)  & 5.70 & 0.549139(63)   \\
${S_{0}}^\ddag$ & 1.00 & 0.58549(54) & 2.25 & 0.586310(60) && --- \\
\cite{Lucini:2001ej} &     &         ---        & 2.25 & 0.586207(29) & 5.70 & 0.549123(56) \\
\end{tabular}
\end{ruledtabular}
\caption{\label{tab:plaquettes}
Expectation values of the plaquette operator $u_p \equiv W(\square)$ in numerical simulations of the various $n$-link actions, estimated from $10^5$ configurations generated on a $8^4$ lattice. For the 0-link action we evaluate the plaquette vev using both \eqref{eq:WilsonLoop} ($\dag$) and \eqref{eq:energy-density} ($\ddag$), whenever possible. We also compare our results with the literature \cite{Lucini:2001ej}.}
\end{table}

For the 0-link models we used the configurations of $Q,R$ variables generated in the simulation of the 1-link model. This is equivalent to treating the unitary link variables as auxiliary to the dynamics of the Gaussian variables. The expectation value of the 0-link plaquette operator \eqref{eq:WilsonLoop} is consistent with the expectation value calculated with the other $n$-link actions (see Table \ref{tab:plaquettes}).

The accurate computation \cite{Amos:1974} of modified Bessel functions in \eqref{eq:0-link:S:SU2} and their ratios in \eqref{eq:energy-density:SU2} and \eqref{eq:effective-link:SU2}, for large $J_l$, is essential to obtain the correct expectation value of 0-link observables in the $SU(2)$ gauge theory. For $SU(3)$, the effective link is constructed numerically with a simple Monte Carlo averaging.
\\

\lettersection{Discussion} The one-link integrals \eqref{eq:group_integral} can ultimately be expressed as power series \cite{Eriksson:1980rq} of the components $Q_{x,\mu\nu}^{ij}$ and $R_{x,\mu\nu}^{ij}$. Therefore, the Gaussian integrals in the 0-link partition function \eqref{eq:0-link:Z} can be solved analytically, term by term, at least in principle.

The Gaussian integration would leave behind residual dynamical degrees of freedom in the form of integer occupation numbers of certain geometrical objects on the lattice, similar to the picture that emerges in the flux representation of the $SU(3)$ spin model \cite{Gattringer:2011gq,*Mercado:2012ue}. Such a representation for the simplest gauge groups, $U(1)$ and $SU(2)$, may actually be constructed explicitly, which we leave for future publications. 

For larger $N$, such representations are much harder to construct. However, we do not exclude the possibility that different HS transformations and/or lattice geometries may lead to simpler and more symmetric expressions for $J_{\mu,x}$, and consequently for the 0-link partition function, which could circumvent such a difficulty.

In practice, these new representations do not bring any clear advantage to the simulation of pure gauge theories: extra Gaussian degrees of freedom require more computational time and they worsen autocorrelations. However, the 1-link and 0-link cases provide suitable representations for the simulation of lattice gauge theories with matter fields.

In fact, it is straightforward to extend the 0-link action \eqref{eq:0-link:S} to include $N_f$ flavours of staggered fermions, by simply generalizing the one-link integrals \eqref{eq:group_integral} with sources of the form $\frac{\beta}{2N} J_{x,\mu} + \sum_{\alpha=1}^{N_f} K_{x,\mu}^\alpha$, where $K_{x,\mu}^{\alpha ij} \propto \psi_x^{\alpha i} {\bar\psi}_{x+\hat\mu}^{\alpha j}$ are $N \times N$ fermionic matrices with pure Grassmann-even components. This is possible because the staggered action is already linear with respect to the link variables. Such an extension is the natural step towards a worldline representation of finite density lattice QCD at finite $\beta$, on which we will elaborate further in future publications.
\\

\begin{acknowledgments}
\lettersection{Acknowledgements} We would like to thank Jo\~{a}o Penedones for many helpful discussions. This work is supported by the Swiss National Science Foundation under grant 200020-149723. HV was supported by Funda\c{c}\~{a}o para a Ci\^{e}ncia e a Tecnologia (Portugal) under grants SFRH/BPD/37949/2007 and CERN/FP/123599/2011. Simulations were performed in the Avalanche cluster (Porto) and in the Brutus cluster (ETH).
\end{acknowledgments}

\bibliography{references}

\end{document}